\documentclass[a4paper,11pt,english]{article}
\pdfoutput=1

\usepackage[separate-uncertainty]{siunitx}
\usepackage{jinstpub}
\usepackage{booktabs}
\usepackage{tikz}

\begin{document}

\title{PANCAKE: a large-diameter cryogenic test platform with a flat floor for next generation multi-tonne liquid xenon detectors}

\author{Adam~Brown,}
\author{Horst~Fischer,}
\author{Robin~Glade-Beucke,}
\author{Jaron~Grigat,}
\author{Fabian~Kuger,}
\author{Sebastian~Lindemann,}
\author{Tiffany~Luce,}
\author{Darryl~Masson,}
\author{Julia~Müller,}
\author{Jens~Reininghaus,}
\author{Marc~Schumann,}
\author{Andrew~Stevens,}
\author{Florian~Tönnies,}
\author[1]{Francesco~Toschi%
\note{Now at Institute for Astroparticle Physics, Karlsruhe Institute of Technology, 76021 Karlsruhe, Germany.}}
\emailAdd{adam.brown@physik.uni-freiburg.de}
\emailAdd{sebastian.lindemann@physik.uni-freiburg.de}
\emailAdd{julia.mueller@physik.uni-freiburg.de}
\affiliation{Physikalisches Institut, Universität Freiburg, 79104 Freiburg, Germany}

\abstract{%
The PANCAKE facility is the world's largest liquid xenon test platform.
Inside its cryostat with an internal diameter of \qty{2.75}{m}, components for the next generation of liquid xenon experiments, such as DARWIN or XLZD, will be tested at their full scale.
This is essential to ensure their successful operation.
This work describes the facility, including its cryostat, cooling systems, xenon handling infrastructure, and its monitoring and instrumentation.
The inner vessel has a flat floor, which allows the full diameter to be used with a modest amount of xenon.
This is a novel approach for such a large cryostat and is of interest for future large-scale experiments, where a standard torispherical head would require tonnes of additional xenon.
Our current xenon inventory of \qty{400}{kg} allows a liquid depth of about \qty{2}{cm} in the inner cryostat vessel.
We also describe the commissioning of the facility, which is now ready for component testing.
}


\maketitle

\section{Introduction}
\label{sec:introduction}

Cryogenic xenon-based detectors have world leading discovery potential to rare particle physics phenomena, such as interactions of WIMP dark matter and neutrinoless double beta decay~\cite{XENON:2023cxc, LZ:2022lsv, PandaX-4T:2021bab, NEXT:2023daz, EXO-200:2019rkq}. 
The next generation of such experiments will capitalise on their success and extend their science reach.
The required exposure and background level will only be possible with a larger detector~\cite{Schumann:2015cpa, nEXO:2017nam, NEXT:2020amj}.
Prominent examples of proposed future dark matter experiments are DARWIN~\cite{DARWIN:2016hyl} and XLZD~\cite{Aalbers:2022dzr}.
The thermal cycling inherent to such experiments introduces challenges related to contraction and mechanical stability of components, worsening with increasing size.
One specific example is the electrodes, which are needed to produce the fields to drift electrons through the liquid xenon and extract them into the gas.
These electrodes will require higher voltages in order to achieve similar performance to current experiments. 
Phenomena seen in current experiments such as spurious emission of light and charge, wire sagging, radiogenic backgrounds and electrostatic breakdown will be problematic with larger electrodes. 
The PANCAKE facility provides a platform for testing full-scale electrodes and other large components which are needed in dual-phase time projection chambers (TPCs).
Its vacuum-insulated stainless-steel cryostat with an internal diameter of \qty{2.75}{m} can reproduce the conditions found in such experiments.
This diameter is based on the proposed design of the DARWIN experiment~\cite{DARWIN:2016hyl}, with a \qty{40}{t} liquid xenon target.
The complementary Xenoscope facility, with a \qty{3.1}{m} tall cryostat, acts as a full-scale vertical demonstrator for future liquid xenon experiments~\cite{Baudis:2021ipf}.

A unique feature of the PANCAKE facility's inner cryostat vessel's design is a flat floor, which saves the tonnes of liquid xenon which would otherwise be needed to fill a standard torispherical pressure-vessel head.
We demonstrate the use of an open-topped vessel within the inner cryostat vessel, allowing the diameter over which liquid xenon is filled to be adapted to the requirements of a particular test.
We are not aware of previous examples of such a vessel being filled with liquid xenon.
Cooling is provided by liquid nitrogen, with a thermosyphon providing adjustable power.
Details of the cryostat and cooling are in \autoref{sec:cryogenics}.
Xenon storage and purification systems, described in \autoref{sec:gassystem}, allow operation with purified xenon.
\autoref{sec:slowcontrol} outlines the SCADA system used to continuously monitor and control the facility. \autoref{sec:commissioning} describes the commissioning of the facility with \qty{300}{kg} of xenon.

\section{Cryostat and cooling}
\label{sec:cryogenics}

\begin{figure}
    \centering
    \includegraphics[width=225.9pt]{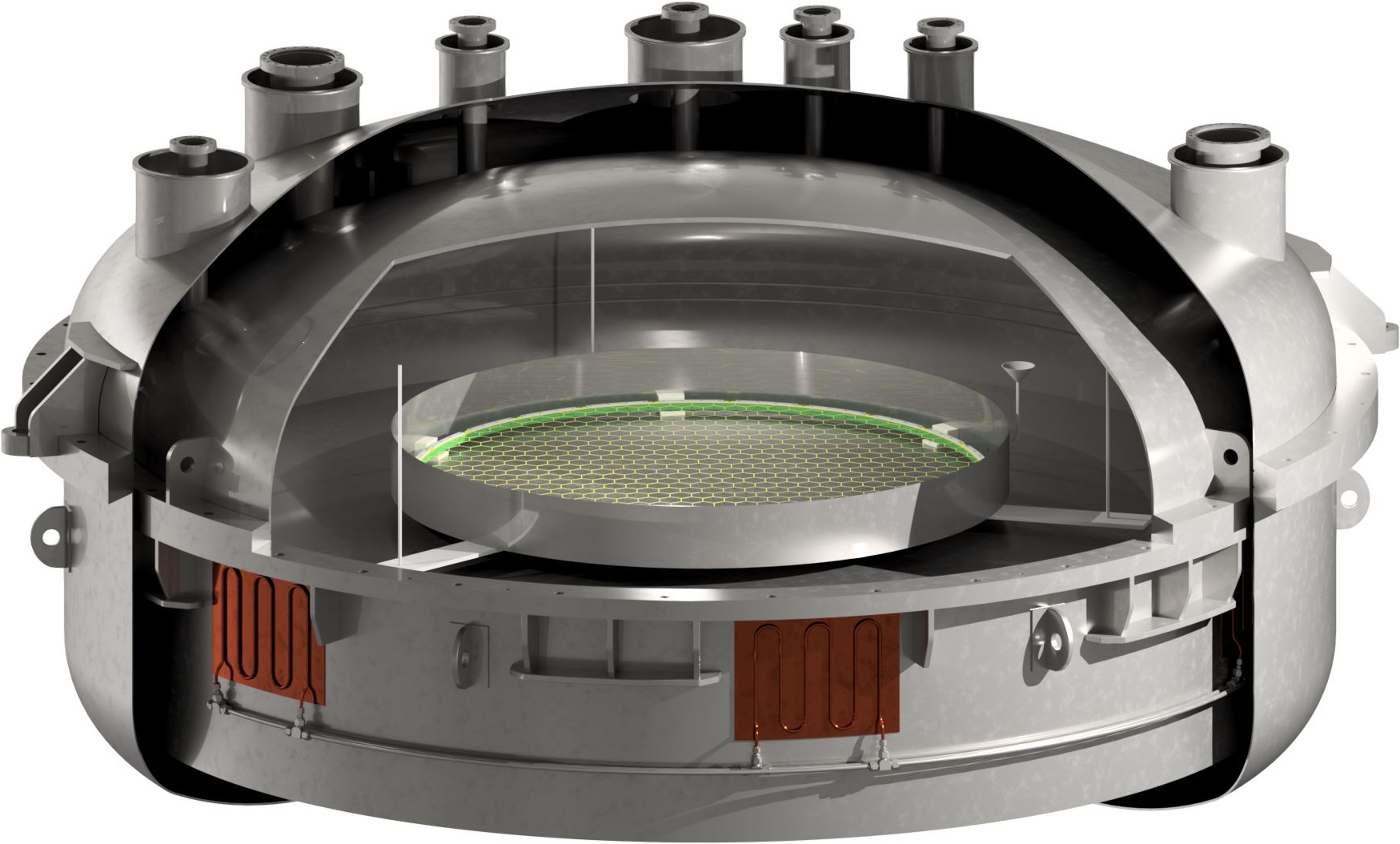}
    \hfill
    \includegraphics[width=190pt]{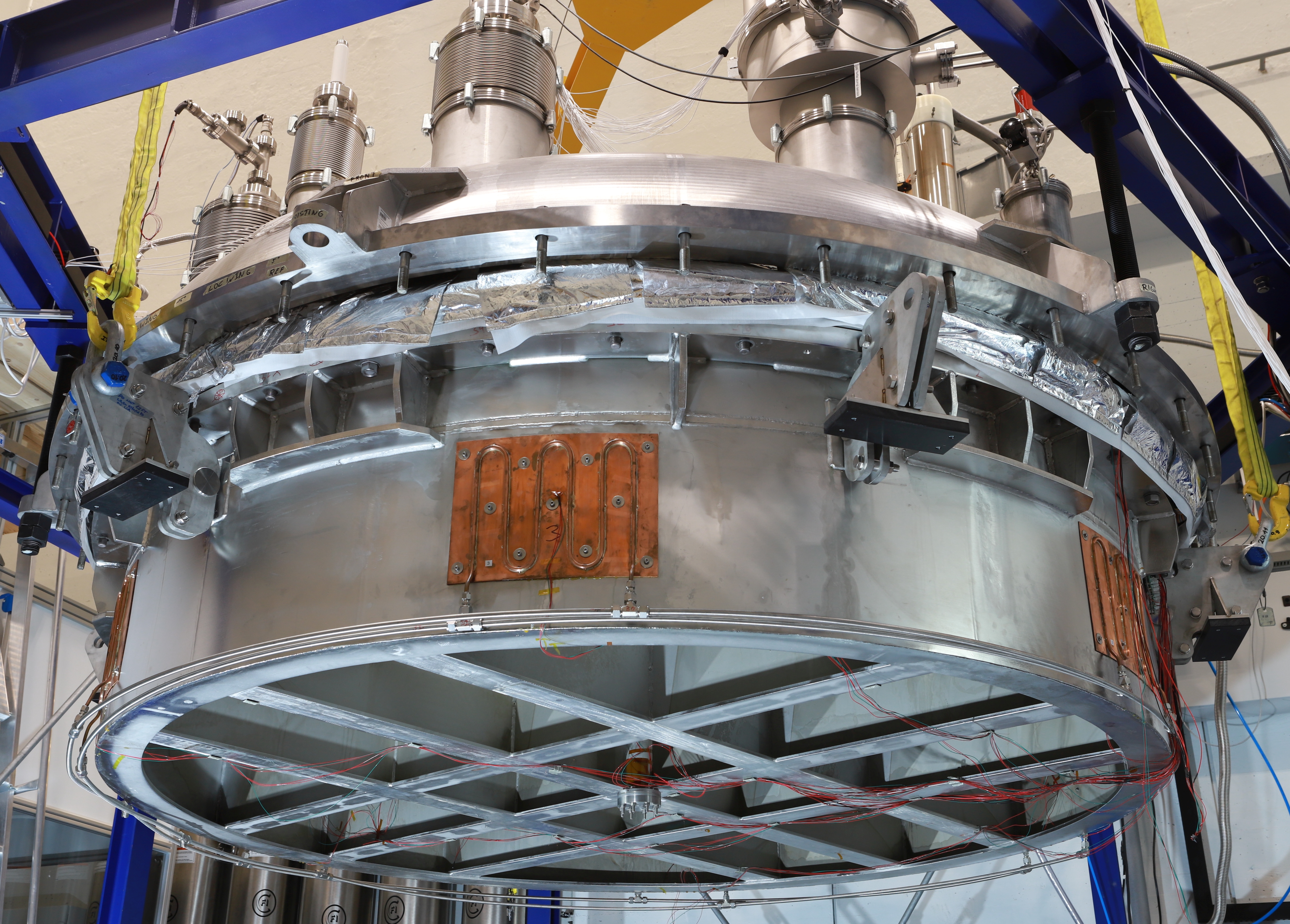}
    \caption[Pancake overview]{\textbf{Left}: a cut-away rendering of the PANCAKE facility's vacuum-insulated cryostat with a \qty{2.75}{m} inner diameter. On top of the cryostat are the CF-63 and CF-150 inner-vessel ports and their corresponding outer-vessel feedthroughs. The external cooling pads on the outside of the inner vessel are visible. For illustration, a test electrode (yellow hexagonal mesh) is placed inside the \qty{1.48}{m} open-topped vessel described in \autoref{sec:commissioning}. The superinsulation is omitted. \textbf{Right}: photo of the inner cryostat vessel suspended from the upper part of the outer vessel. The additional stiffening structure used to support the flat floor and the external cooling pads are visible.}
    \label{fig:cryostat}
\end{figure}

The PANCAKE facility incorporates a vacuum-insulated cryostat, comprising two stainless steel vessels, as seen in \autoref{fig:cryostat}.
The cryostat is suspended from a support structure constructed from steel beams.
Two liquid-nitrogen based systems can regulate the inner cryostat vessel's temperature between around \qty{-200}{\celsius} and room temperature.

\subsection{Cryostat} \label{sec:cryostat}
At the core of PANCAKE is the inner cryostat vessel, which is split into an upper section with a torispherical head and a lower section with a flat floor.
The cryostat is constructed from grade EN~1.4301 stainless steel,  \qty{5}{mm} thick except for the floor, which is \qty{15}{mm}. 
It has an internal diameter of \qty{2.75}{m} and \qty{180}{mm} height between the floor and flange.
Over this diameter, \qty{17}{kg} of liquid xenon is needed per \qty{1}{mm} of filling depth at an absolute pressure of \qty{2}{bar}.
The internal volume is \qty{3.6}{m^3}, including the upper section.

An additional stiffening structure is welded to the bottom of the vessel to ensure the floor remains flat when the vessel is pressurised.
A \qty{5}{mm} thick cylindrical plate extends the wall of the inner vessel downwards by 425 mm and encloses two perpendicular sets of five \qty{5}{mm} stiffening plates.
The height of the three innermost plates is \qty{425}{mm}, while the outer two are \qty{250}{mm}.
This restricts the deflection at the centre of the floor to \qty{1.3}{mm} under a relative pressure of \qty{2}{bar}, meaning \qty{12}{kg} of additional liquid xenon is needed to cover the floor.
This deflection was estimated 
using finite-element method simulations, and confirmed in a pressure test within 5\%.
The floor was machined after welding to further ensure its flatness, after which the inner surface of the whole vessel was electropolished.
Because of the thick floor and the stiffening structure, the bottom part of the vessel weighs \qty{1.9}{t}, compared to \qty{1.2}{t} for the top part.
Without a flat floor, tonnes of xenon would be needed to submerge test components over the full diameter of a standard pressure vessel.

The inner vessel is sealed with a \qty{2}{mm} oxygen-free high conductivity (OFHC) copper wire, soldered into a ring using Sn96Ag4 lead-free solder.
The wire is positioned between two \qty{5}{cm}~thick and \qty{10}{cm}~wide flanges.
The flanges are connected using 36~M22 grade~\mbox{A4-80} stainless steel bolts, resulting in a leak rate better than \qty{3e-4}{mbar.L/s}, as discussed in \autoref{sec:commissioning}.

RUAG Space COOLCAT 2 NW superinsulation is used to reduce the radiative heat load on the inner vessel to below~\qty{10}{W} at \qty{-100}{\celsius}.
Each of its thirty layers is made of \qty{12}{\um} polyester foil, aluminised on both sides and perforated to allow efficient pumping.

Three stainless steel supports connect the top parts of the inner and outer vessels.
These supports are designed to withstand the atmospheric pressure on the inner-vessel ports equivalent to \qty{4.1}{t}.
These are attached to the inner and outer vessels on pivots, allowing the inner vessel to shrink during cooling.
Their design minimises the thermal conductance between the inner and the outer vessel, resulting in a simulated heat flow of \qty{6.5}{W} across all three supports at a \qty{125}{K} temperature difference.

The outer vessel is also constructed from EN~1.4301 stainless steel and split into two sections with torispherical heads.
It has an overall external height of \qty{1.83}{m} and a diameter of \qty{3.21}{m}, with a mass of \qty{1.0}{t} for the upper and \qty{1.1}{t} for the lower section.
The upper and lower section are sealed using a Viton seal between two flanges with the same cross-section as the inner vessel.
These are connected using 36 M20 stainless steel bolts.

Just Vacuum GmbH designed and manufactured the cryostat and certified the inner vessel as a pressure vessel up to \qty{4}{bar}~relative, according to EU directive 2014/68/EU (PED).

\begin{figure}
    \centering
    \includegraphics[width=237.5pt]{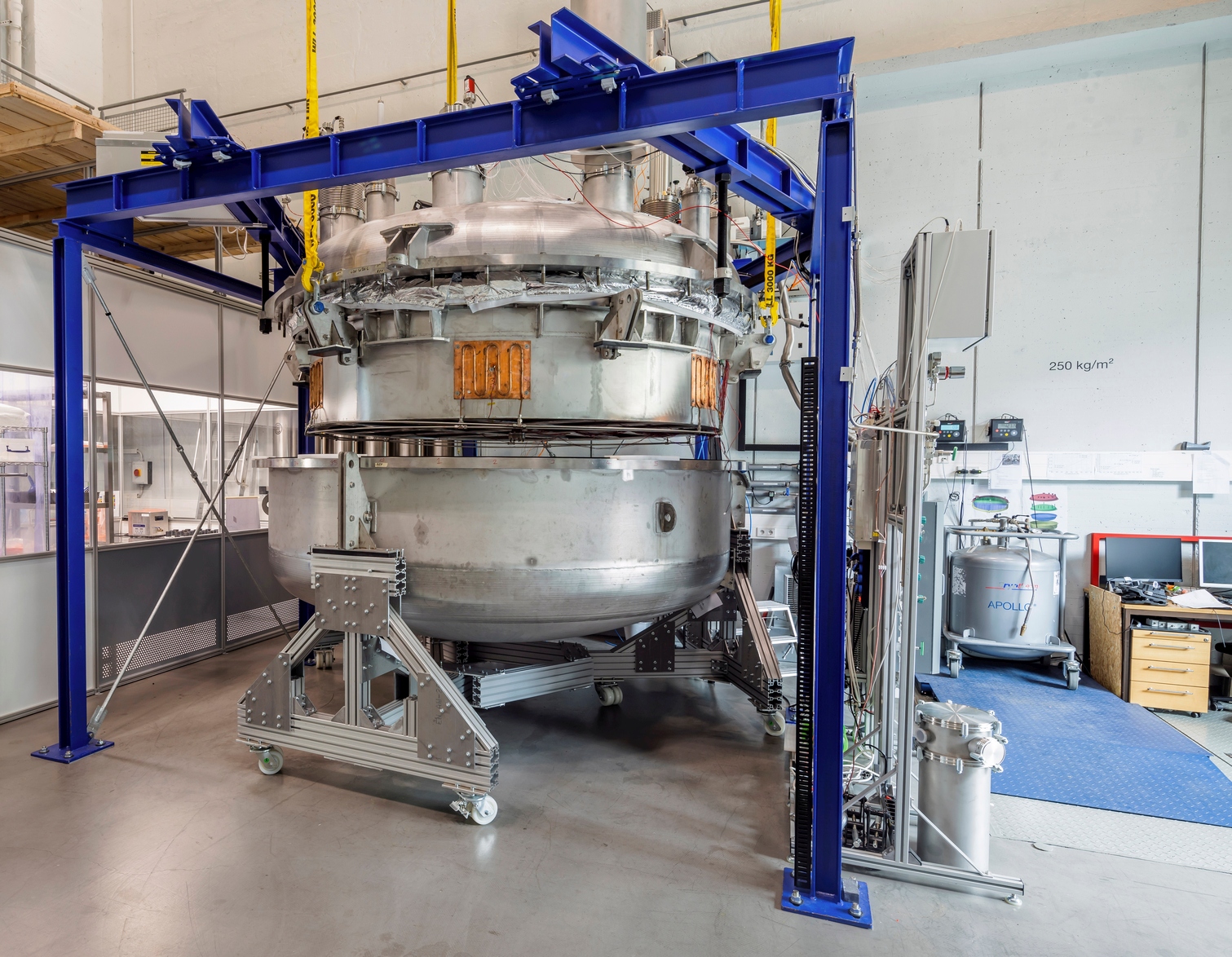}
    \hfill
    \begin{tikzpicture}
    \draw (0, 0) node[inner sep=0] {\includegraphics[width=178.5pt]{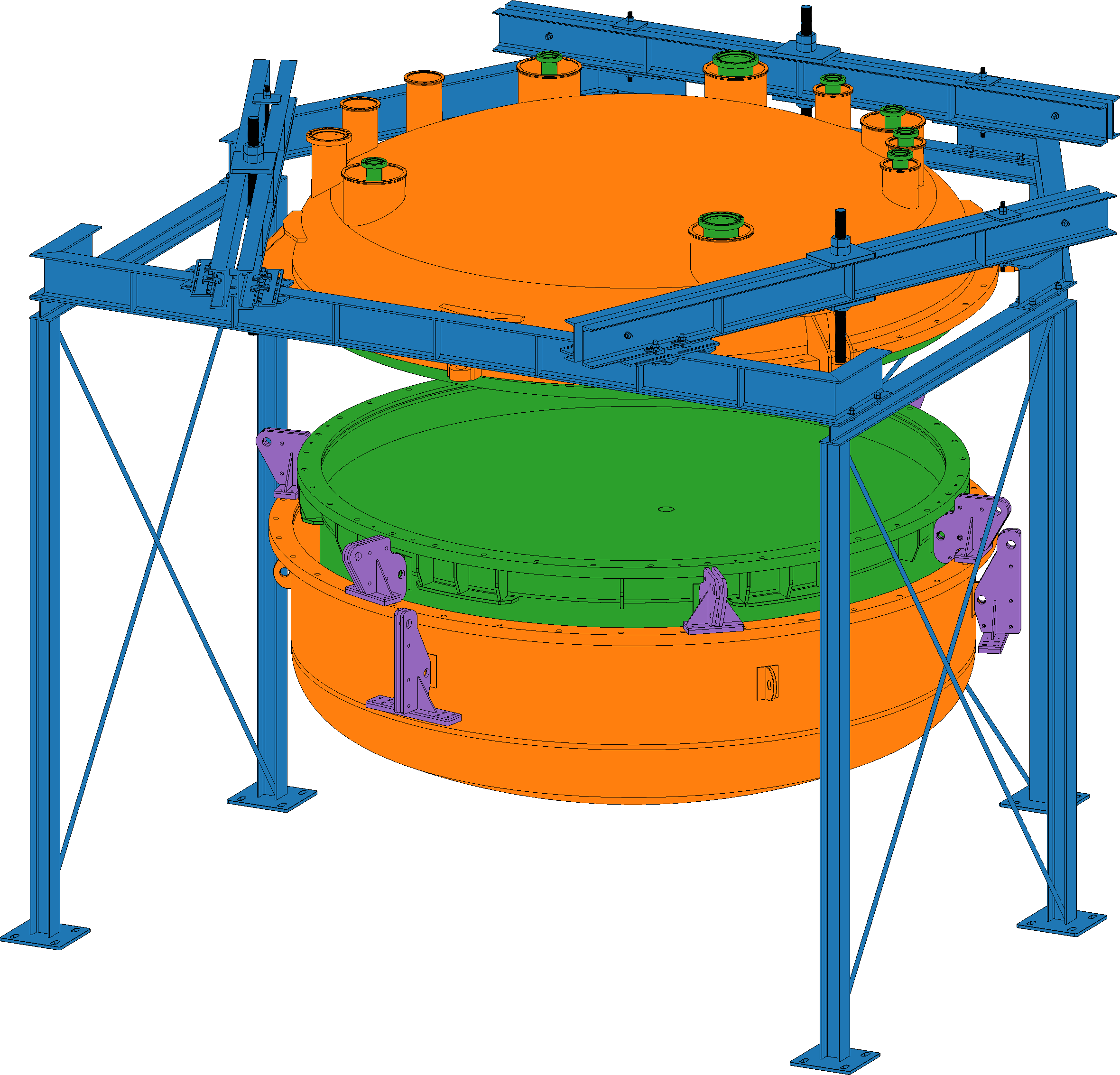}};
    \draw (-1.1, 0.2) node[anchor=south west] {\small Lower inner part};
    \draw (-1.1, -0.1) node[anchor=south west] {\small Flat floor};
    \draw (0.2, -0.9) node[anchor=north] {\small Lower outer part};
    \draw (0.5, 1.8) node[anchor=south] {\small Top outer part};
    \draw (1.5, 3.0) node[anchor=south] {\small M45 rod};
    \draw[line width=1.5pt] (-1.5,-2.5) -- (-0.3,-2.5);
    \draw[line width=1.5pt] (-1.5,-2.575) -- (-1.5,-2.425);
    \draw[line width=1.5pt] (-0.3,-2.575) -- (-0.3,-2.425);
    \draw (-0.9, -2.5) node[anchor=south] {\small 1 m};
    
    \end{tikzpicture}
    \caption[Pancake Overview]{Photo (left) and drawing (right) of PANCAKE in its support structure, the lower parts are opened downwards.
    The top outer part (orange) is fixed to the support structure (blue) by three M45 rods (black).
    The inner vessel (green) hangs within the top outer part.
    Additional stainless steel extensions (purple) are used to attach the bridge crane in order to lift the inner vessel.
    }
    \label{fig:pancake_ov}
\end{figure}
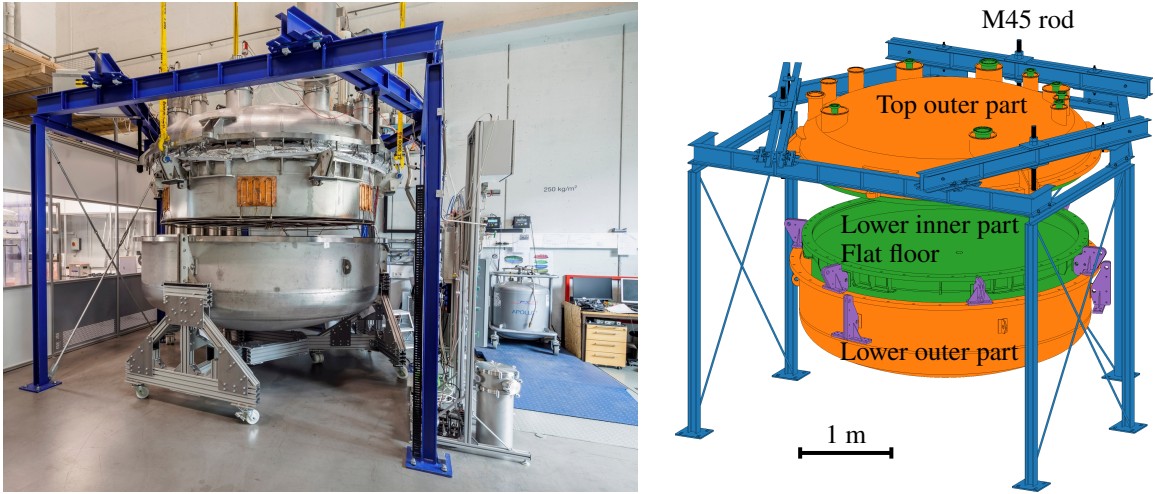

Two CF-150 and six CF-63 ports are available on the top of the inner vessel, passing through the isolation vacuum, for instrumentation and accessories.
The outer vessel itself has one additional CF-160 and two ISO-K-150 ports.
There is also a central CF-63 port at the bottom of the inner vessel and an accompanying ISO-K-150 on the outer vessel that could be used in the future for gravity-assisted liquid xenon extraction.

The inner and outer vessel can each be evacuated using independent Leybold Turbovac~450i tubomolecular pumps installed on the CF-150 flanges.
These are each backed up by a rotary vane pump: one Pfeiffer Pascal 2005SD and one Oerlikon Leybold AMM 71Z BA4, allowing evacuation to \qty{e-4}{mbar} in approximately one day.
The inner vessel's evacuation port can be closed with an MDC AV-600M CF-150 angle valve, between the pump and the vessel itself.

\subsection{Support frame and access}
The outer cryostat vessel hangs from a support structure on three \qty{45}{mm} threaded rods, with \qty{4.5}{mm} pitch, seen in \autoref{fig:pancake_ov}.
The cryostat can be levelled by adjusting these three rods.
The support frame distributes the load over six points on the floor and is fixed at two points to the wall for added stability.

The cryostat is opened by removing the lower section of each vessel.
The upper sections of both vessels are left in place, allowing components to be left installed on their ports.
Each lower section in turn is lowered using a bridge crane.
The two vessels are stored inside one another on a custom cart, which can then be rolled out through a large opening at the front of the support frame.

Two Envirco MAC 10XL FF25A-001 fan filter units can be mounted within an acrylic roof at the top of the support frame.
Together with removable PVC strip curtains around the frame, these can provide a low-dust environment to install components in the cryostat.

\subsection{Cooling}
\label{ssec:cooling}
\begin{figure}
    \centering
    \includegraphics{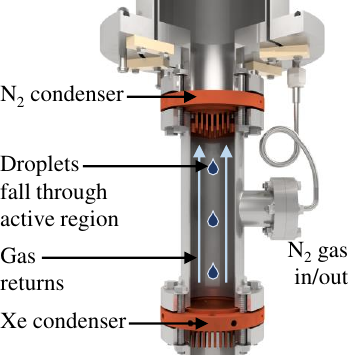}
    \hfill
    \includegraphics{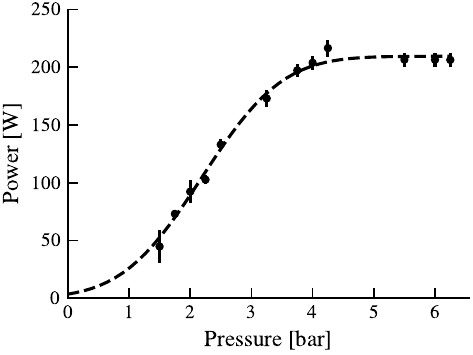}
    \caption[Thermosyphon working principle and results]{\textbf{Left}: cut-away view of the thermosyphon, showing the reservoir at the top, the active region in the centre, and the xenon volume at the bottom. Nitrogen gas can be added to or removed from the active region through the pipe seen on the right of the image. \textbf{Right}: cooling power of the thermosyphon as a function of active region pressure at a typical operating temperature of \qty{-100}{\celsius} at the xenon condenser.}
    \label{fig:TS-workingprinciple}
\end{figure}

In stable operation, temperature is maintained by liquefying xenon within the inner vessel with a thermosyphon \cite{Bradley:2012fsa}, seen in \autoref{fig:TS-workingprinciple} (left).
The thermosyphon is based around an active region with an internal diameter of \qty{66}{mm} and a height of \qty{210}{mm}.
At the top, a condenser separates it from a \qty{17}{L} liquid nitrogen reservoir.
Gaseous nitrogen in the active region is condensed here and the droplets produced fall to the bottom.
They then evaporate and in doing so cool a second, lower condenser, which separates the active region from the inner cryostat vessel.
This second condenser is used to liquefy xenon.
Both condensers are made from oxygen-free high-conductivity copper, used in place of the gasket between two CF-63 flanges.

The cooling power of the thermosyphon can be adjusted by varying the pressure of gaseous nitrogen in the active region.
A maximum power of about \qty{200}{W} is achieved at \qty{5}{bar}, as shown in \autoref{fig:TS-workingprinciple} (right), measured using the heating power needed to stabilise the temperature with vacuum below the lower condenser.
Five \qty{150}{W} heating cartridges and one PT100 are installed in the centre of the lower condenser. 
A Cryocon 26C regulates the temperature of the condenser by providing up to \qty{100}{W} of total heating power to four of the heating cartridges.
The fifth can be manually powered and is used when the maximum power available from the Cryocon 26C is insufficient.
The upper condenser is equipped with two \qty{150}{W} heating cartridges to remove water which enters the liquid nitrogen reservoir.

The liquid nitrogen reservoir is a vacuum-insulated stainless-steel cylinder.
A polyamide lid with extruded polystyrene insulation on the inside minimises water and heat ingress.
The reservoir is automatically refilled from a pressurised nitrogen dewar through an ASCO 263LT solenoid valve, based on capacitive level monitoring.
At a typical cooling power of roughly \qty{150}{W}, the reservoir is refilled every two hours.
An exhaust pipe through the lid directs liquid nitrogen away from critical parts of the facility in case of overfilling.

It is difficult to directly measure the heat load on the inner vessel when cold.
During stable operations, the thermosyphon was operated with an active region pressure of \qty{2.84(0.01)}{bar} and a heater power of \qty{60}{W}, corresponding to an overall cooling power of \qty{94(11)}{W}.
This was measured with \qty{300}{\kg} of liquid xenon in the open-topped vessel (see \autoref{sec:commissioning}), a xenon pressure of \qty{1.4}{bar}, a temperature of \qty{-76}{\celsius} at the inner vessel's flange and with roughly \qty{10}{W} electrical power supplied to the cameras (see \autoref{sec:cameras}).
The heat load was directly measured based on the rate of the inner vessel warming up to be \qty{55(14)}{W} at \qty{-60}{\celsius}, once the temperature had equalised across the evacuated inner vessel.
As well as the \qty{16}{\celsius} difference in temperature between the two measurements, thermal gradients in the inner vessel may contribute to the disparity between the heat load and the cooling power.

The expected conductive heat load with liquid xenon in the cryostat is about \qty{20}{W} through the cryostat ports and the connections between the two cryostat vessels, assuming a uniform inner vessel temperature of \qty{-76}{\celsius}.
The radiative heat load is expected to be below \qty{10}{W}.
With an insulation vacuum better than \qty{e-3}{mbar}, convective heat transfer is smaller than around \qty{3}{W}.
The discrepancy between measurements and expectations can be explained by additional heat intake channels, for which quantitative estimations are unavailable.
These include transport by the gaseous xenon inside the inner vessel and radiation absorbed by accessories with no superinsulation, such as the thermosyphon and heat exchanger.

A second liquid-nitrogen based cooling system accelerates the initial cooling of the inner cryostat vessel.
This consists of six \qty{40}{cm} by \qty{30}{cm} copper pads which are bolted directly to the wall of the inner vessel's stiffening structure, as seen in \autoref{fig:cryostat}.
Apiezon~N thermal paste between the stainless steel wall and the cooling pads improves the thermal contact.
A \qty{1.8}{m} long, \qty{8}{mm}~outer-diameter copper pipe in a serpentine pattern is brazed onto each pad.
Liquid nitrogen flowing through the pipe provides up to around \qty{100}{W} of cooling per pad.
The inlet and outlet of the six pads are connected in parallel to the feedthrough at the bottom of the outer cryostat vessel. 
This external cooling system is controlled by an ASCO~263LT solenoid valve at the outlet.
The duty cycle for this valve is typically chosen such that remaining liquid nitrogen is just beginning to reach the outlet, meaning the maximum available power in the phase transition is exploited.
With both cooling systems running, it takes about two weeks to cool the inner cryostat vessel from room temperature to the \qty{-100}{\celsius} operating temperature.

\section{Xenon gas handling}
\label{sec:gassystem}

\begin{figure}
    \centering
    \includegraphics[width = \textwidth]{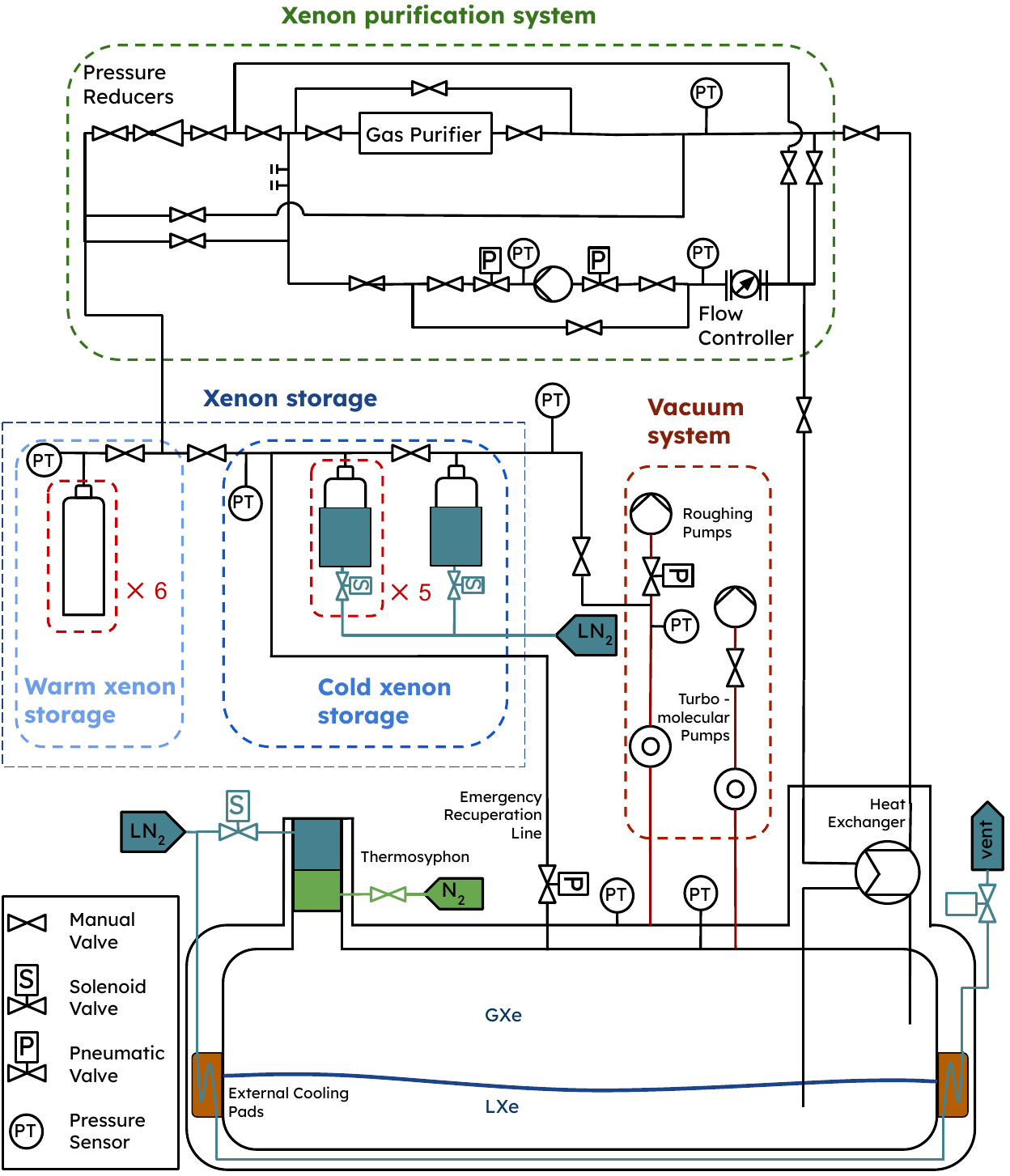}
    \caption{Simplified piping and instrumentation diagram. Xenon lines are black, gaseous (N$_2$) and liquid (LN$_2$) nitrogen lines blue, and vacuum lines red. Some valves are omitted on this diagram to improve clarity. The liquid nitrogen distribution system supplying the xenon storage and cooling systems is not shown here.}
    \label{fig:PnID_simplified}
\end{figure}

Two systems are used to handle the xenon, which are seen together with their connections in \autoref{fig:PnID_simplified}.
A purification system removes electronegative impurities such as oxygen and water from the xenon.
Between operation cycles, xenon is contained in a storage system.

\subsection{Storage system}
\label{sec:storagesystem}
When not in use, the xenon is stored in twelve gas bottles, split between two storage units.
Six \qty{50}{L} bottles form the cold xenon storage, which can be used to recover xenon from the cryostat. The remaining six (two \qty{40}{L} and four \qty{50}{L}) bottles provide additional storage capacity in the warm xenon storage.
To fill the cryostat, xenon from any of the twelve storage bottles can be fed to the inner cryostat vessel via the purification system, described below.

The cold xenon storage bottles are made of aluminium and suspended from MecSense TS/3kN load cells to monitor their xenon content.
Each bottle can be individually cooled by filling liquid nitrogen into an open dewar surrounding the bottle.
The liquid nitrogen level in each dewar is monitored by a capacitive levelmeter and a Gram Xtrem F1-150 floor scale, and refilled through automatically-controlled ASCO 263LT solenoid valves. 

Once cooled to liquid nitrogen temperature, the vapour pressure in the bottles drops below \qty{1}{mbar}.
This results in gaseous xenon flowing from the cryostat to the cold storage, where it freezes in the bottles.
After recuperation, the bottles are warmed up to room temperature.
The warm xenon storage provides additional capacity and its content is monitored by a Gram XTiger floor scale.

Recuperation from the inner vessel is possible via the purification system with flow measurement.
An independent automated path controlled by a Swagelok SS-8BK-VCR-1C normally-closed pneumatic valve allows for higher recuperation flows in emergency cases.
Recuperation from the isolation vacuum is also possible, allowing xenon which leaks through the inner vessel's seal following a failure to be recovered.
To do this, a Hositrad VAB25KF pneumatic valve isolates the outer vessel's backing pump, while an additional Swagelok SS-8BK-VCR-1C valve redirects the turbomolecular pump's outlet to one of the cold xenon storage bottles.

\subsection{Purification system}

The core function of the purification system is to remove electronegative impurities, which constantly outgas from surfaces in contact with xenon.
It also acts as an interface between the storage system and the cryostat.

For continuous purification during operation, liquid xenon is extracted from the inner vessel through a heat-exchange system consisting of a Kelvion GVH 500H-80 plate heat exchanger, where xenon is evaporated, and a GVH 300H-40, which warms it to almost room temperature.
A KNF N143 SP.12E diaphragm pump circulates the xenon through the purification system.
A second, redundant diaphragm in the pump avoids xenon loss in case the primary diaphragm ruptures.
Simultaneous failure of both diaphragms could be detected by monitoring the inlet and outlet pressures, allowing the pump to be isolated remotely using two Swagelok SS-8BK-VCR-1O normally-open pneumatic valves.

The flow is controlled by throttling the pump inlet or outlet gas stream and monitored using a Teledyne Hastings HFC-D-303-B flow controller.
A bypass loop allows xenon from the pump outlet to be directed back to the inlet, reducing the flow through the gas system while minimising the pressure gain and wear on the pump.
A SAES Mono Torr PS4-MT15-R rare gas purifier reduces the concentration of impurities in the xenon to below \qty{1}{ppb} at its outlet for flows up to \qty{9}{slpm}.
Four spare VCR ports could allow the injection of gaseous calibration sources and the measurement of xenon purity.
We plan to use the Tiger Optics HALO+ monitor which was previously used by XENON1T to measure the water content of the xenon~\cite{XENON:2017lvq}.
More sophisticated purity monitors could also be used in the future.
Returning gas from the purification system passes through the heat exchanger in reverse, where it is cooled and then condensed.
At the \qty{15}{slpm} flow we have achieved, our inventory of \qty{400}{kg} of xenon can be passed through the purifier once every 3.1 days.

All piping and valves in the purification system use 1/2" VCR face-seal metal-gasket components.
All but one valves in the low-pressure part of the purification system are Swagelok DF-series diaphragm-sealed valves. One Swagelok SS-8BG-VCR bellows-sealed valve provides a degree of manual flow control.
Wherever high pressures from the storage system may reach the purification system, Swagelok 6LVV-DPHFR4-P valves are used.
This results in a leak rate measured to be better than \qty{e-10}{mbar.L/s} per fitting.

To fill the cryostat, xenon from the storage system enters the purification system via two Linde Redline S200 pressure reducers in series.
The second reducer is redundant in normal operation and protects the purification system and cryostat from excessive pressure in case the first fails open, for example freezing due to prolonged high xenon flow.
The gas passes through the flow controller and can then optionally be cleaned by the rare gas purifier before being injected into the cryostat through the heat exchanger.
Xenon can be recuperated from either the liquid or gas phase in the cryostat.
Again, xenon can be passed through the purifier to remove impurities.
The flow controller can be bypassed if necessary to achieve higher flows.
Gravity-assisted liquid xenon extraction or liquid xenon purification could be possible using the cryostat's bottom port, but is not yet implemented. 

\section{Facility monitoring and instrumentation}
\label{sec:slowcontrol}
Reliable monitoring is essential to ensure safety and provide inputs for robust data analysis.
The facility is monitored and controlled in real time using a supervisory control and data acquisition (SCADA) system.
Key components of a SCADA system are generally distributed small-scale computers which monitor and control devices, referred to as programmable logic controllers (PLCs), a central facility for configuration, storage and processing, and the required communication infrastructure.

\subsection{SCADA}
\begin{figure}
    \centering
        \includegraphics[width=0.65\textwidth]{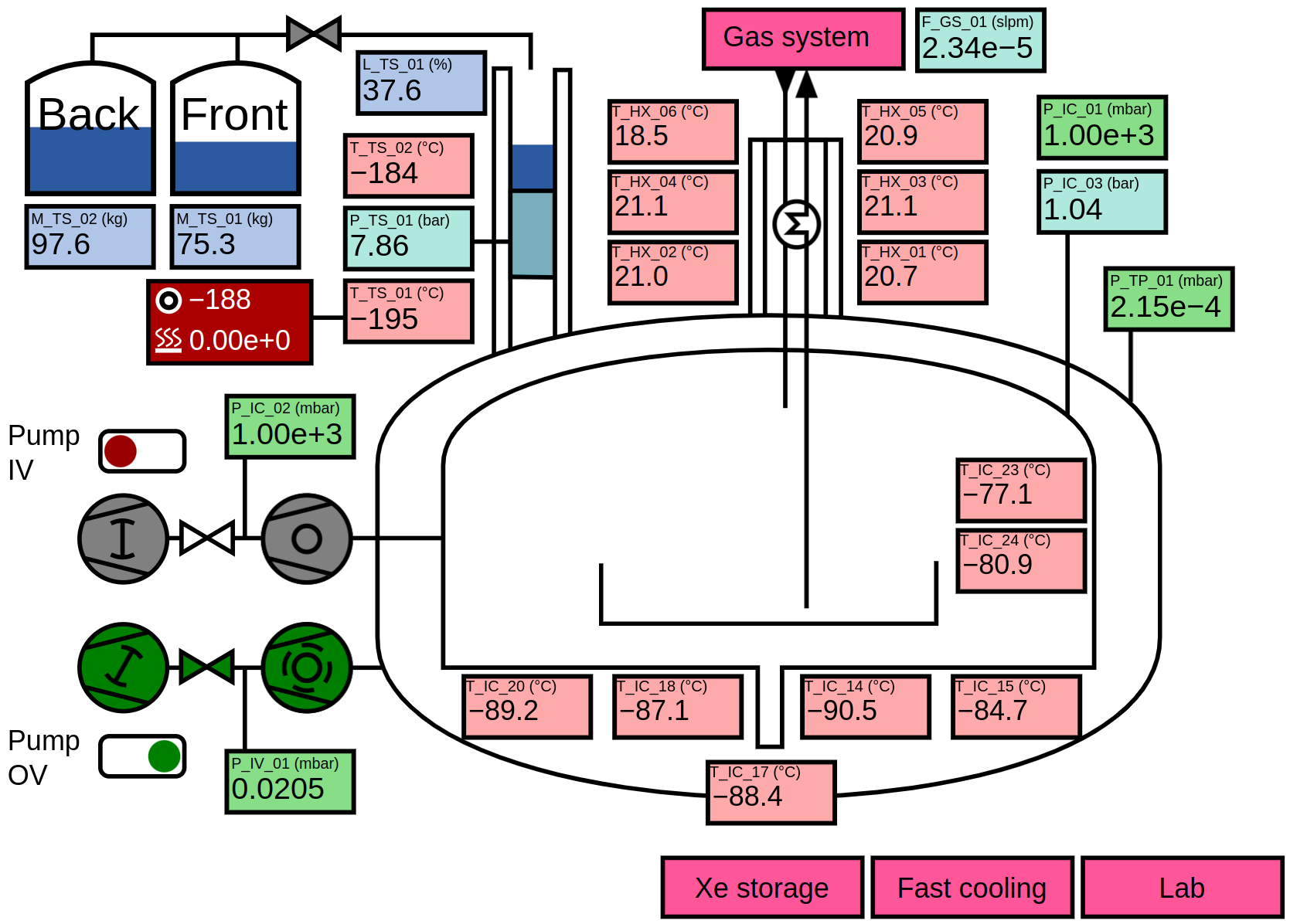} 
	\caption{Interactive web-based control panel for the SCADA system. The coloured boxes display a subset of the latest readings of the various 
	sensors. Clicking on a box shows details about the sensor and a graph of historical readings. Valves and other devices can also be controlled by mouse-click from this display.}
	\label{fig:doberview}       
\end{figure}

For the PANCAKE facility, we use an in-house SCADA system, based on Doberman~\cite{Zappa:2016zsn}.
This lightweight software suite is developed in Python and optimised for small and mid-sized applications.
It has already been deployed for several other applications~\cite{Baur:2022sel, Garcia:2022jdt}.
The central part of the system runs on a Fujitsu server with Ubuntu 20.04.
This server stores the data in the InfluxDB open source (OSS) database, optimised for time-series data.
The configuration settings for the whole system are stored in a MongoDB NoSQL database.
Users can control systems, configure sensors, and view current and historical readings using a web-based interface, seen in~\autoref{fig:doberview}.

The SCADA system collects data from sensors and equipment distributed throughout the facility.
Analogue and digital inputs and outputs are read and controlled by three Revolution Pis acting as PLCs.
These small-scale industrial-grade computers are based on the well-known Raspberry Pi and consist of a DIN-rail-mounted CPU, which can be extended with analogue or digital input and output modules.
Each Revolution Pi is installed in a separate enclosure together with ancillary electronics.
Other sensors and devices can be directly monitored using built-in Ethernet communication protocols.
Ethernet devices can be read by a Doberman instance running either on the central server or one of the Revolution Pis.
Devices which only support serial communication are connected using WaveShare serial-to-Ethernet converters.
Our entire SCADA system is powered via an FSP Clippers RT 3K uninterruptible power supply, which can bridge a power outage of up to an hour.

\subsection{Instrumentation} 
\label{sec:instrumentation}
The vacuum in the inner and outer cryostat vessels is monitored by two Pfeiffer Vacuum PKR~361 full range gauges.
A Pfeiffer Vacuum TPR 280 gauge measures the vacuum between each turbo-molecular pump exhaust and backing pump inlet. 
These sensors as well as the operation of both turbomolecular pumps and their corresponding backing pumps are monitored and controlled by the SCADA system.

A WIKA WU-20 sensor with a range of \qtyrange{0}{7}{bar} monitors the inner cryostat vessel pressure. 
One WIKA WU-20 and two Omega PXM309 sensors, each with a range of \qtyrange{0}{7}{bar}, measure the pressure at three locations in the purification system, shown in \autoref{fig:PnID_simplified}.
An Omega PAA21Y \qtyrange{0}{200}{bar} and two WIKA WU-20 \qtyrange{0}{100}{bar} sensors indicate the pressure in the storage system.
Experience has shown more reliable performance of the WIKA sensors.

Capacitive level meters monitor liquid xenon and liquid nitrogen levels throughout the facility.
Their capacitance depends on the liquid level between their electrodes due to the different dielectric constants of gas and liquid.
Tube-in-tube level meters of roughly \qty{1}{m} length are used to monitor the liquid nitrogen level in the thermosyphon reservoir and the six open dewars of the cold xenon storage system.
The dimensions are slightly different for the two use cases but always result in a capacitance range of around \qtyrange{200}{300}{pF}.
Liquid xenon depths in the inner cryostat vessel are measured using shorter, rectangular level meters.
These consist of three \qty{24.5}{mm} by \qty{40}{mm} parallel plates with alternating polarities separated by \qty{1}{mm}, surrounded by a larger copper shield.
Their design is adapted from that used in XENON1T~\cite{XENON:2017lvq} and is described in~\cite{Toschi_thesis}.
Precise dimensions can be found in \autoref{tab:levelmeter}.
Their capacitance varies between about \qty{23}{pF} and \qty{35}{pF}.
All level meters are connected via RG196A/U coaxial cables to a read-out system based on the Smartec Universal Transducer Interface  \cite{Toschi_thesis,Geis_thesis}, which is controlled and read by the SCADA system.
    
\begin{table}
    \centering
    \caption{Dimensions and capacitances of the level meters.}

    \begingroup
    \let\mathtimes=\times
    \let\times=&
    \begin{tabular}{
      l l l
      S[table-format=2.1]
      @{ $\mathtimes$ }
      S[table-format=1.1]
      @{ $\mathtimes$ }
      l
      S[table-format=5.0]
      @{ -- }
      l
      }
           \toprule
           Location & Medium & Type & \multicolumn{3}{c}{Dimensions [mm$^3$]} &  \multicolumn{2}{l}{Capacitance [pF]} \\
        
           \midrule
           Cold storage  & Nitrogen & Tube in tube & 4 .0 \times 5.5 \times 1120 & 200 \times 280  \\
           Thermosyphon  & Nitrogen & Tube in tube & 5.0 \times 6.5 \times 940 & 210  \times 290  \\
           Inner vessel & Xenon & Parallel plate & 24.5 \times 1.0 \times 40 & 23  \times 35  \\ 
           \bottomrule
        \end{tabular}
    \endgroup
\label{tab:levelmeter}
\end{table}

PT100 sensors measure the temperature throughout the cryostat.
Up to 24 PT100 sensors within the inner vessel measure the temperature of the xenon and test components.
Each PT100 is connected using two RG196A/U coaxial cables, with the core and shield of each cable connected in parallel to allow four-wire readout.
Three Wiesemann~\& Theis 57778 devices located outside the cryostat read eight sensors each.
A further 24 PT100 sensors connected in three-wire mode are positioned within the outer cryostat vessel to monitor the temperature of the inner vessel itself, while another eight sensors measure the temperature throughout the heat exchanger.
These are read out in groups of eight by four Elzet 80 PT100MUX8 multiplexing transducers, providing a \qty{4}{mA} to \qty{20}{mA} output which is read by a Revolution Pi.

\subsection{Cameras}
\label{sec:cameras}
\begin{figure}
	\centering
	
	\begin{tikzpicture}
        \centering
		\tikzstyle{block} = [draw=none,fill=none,text=white];
		\node[anchor=south west,inner sep=0] at (0,0) {\includegraphics[width = 300pt]{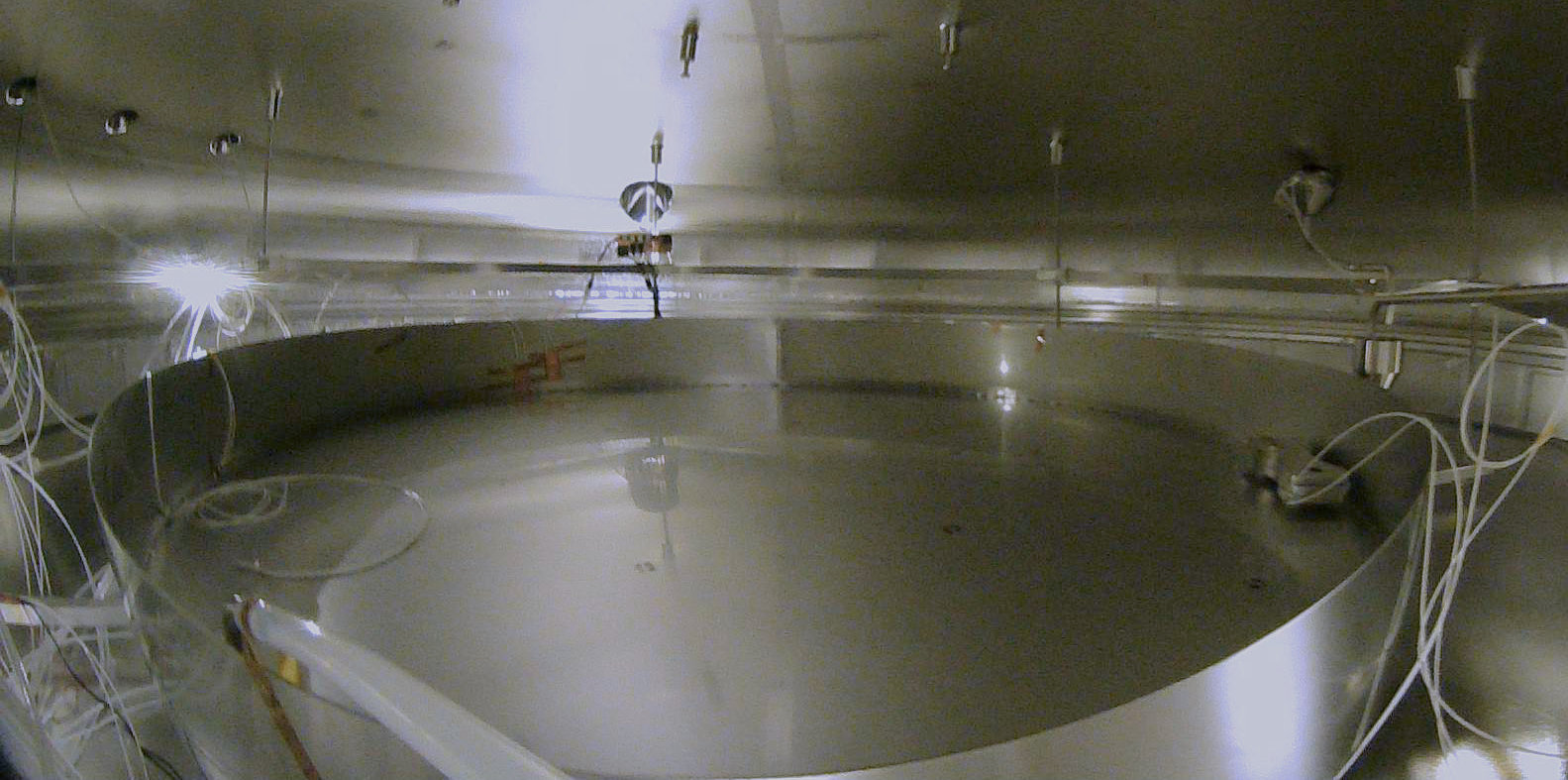}
        };
		\node[block] at (4.5,1) {Open-topped vessel};
        \node[block] at (8,4.5) {Upper part of inner vessel};
        \draw[white, thick] (8.3,1.4) rectangle (9.3,2.5);
        \node[block] at (8.8,1.1) {Test block};
	\end{tikzpicture}
 
    \vspace{10pt}
    \begin{tikzpicture}
        \centering
		\tikzstyle{block} = [draw=none,fill=none,text=white];
		\node[anchor=south west,inner sep=0] at (0,0) {
        \includegraphics[width = 300pt]{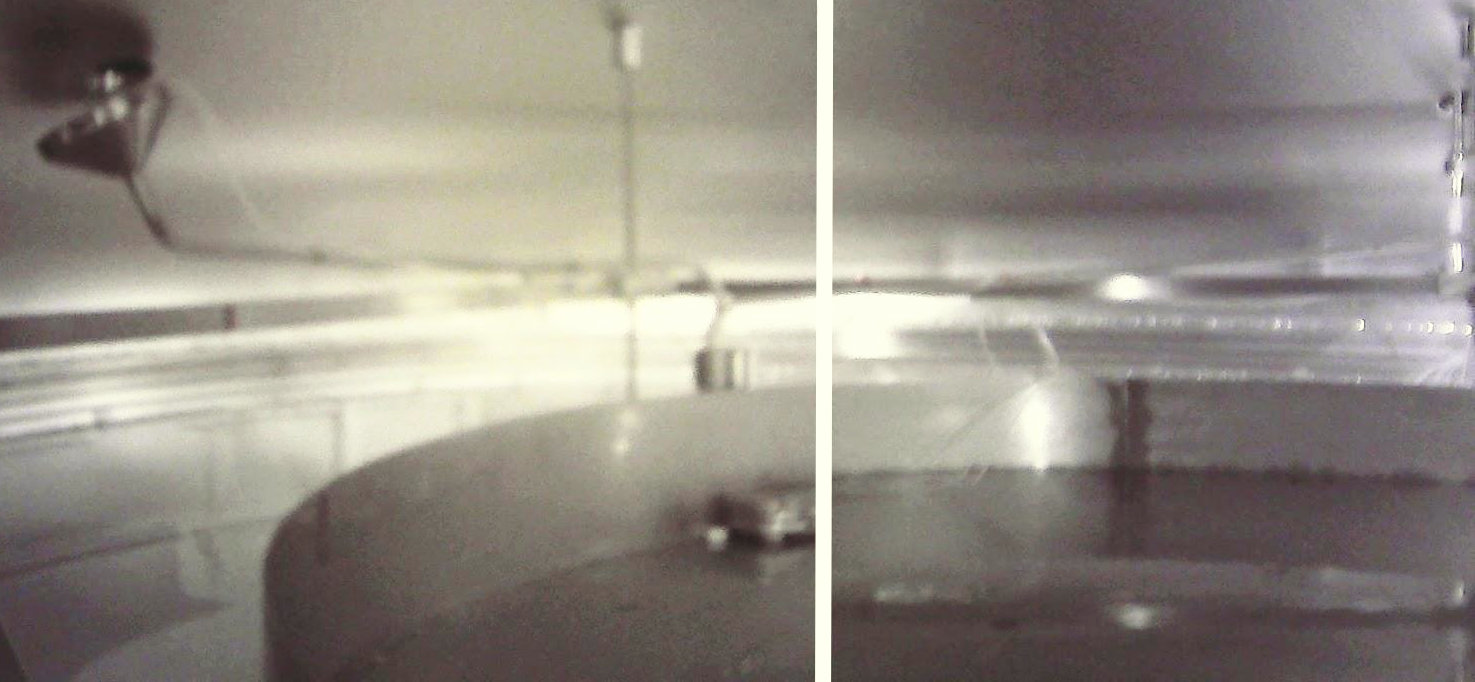}};       
       
        \node[block] at (5.2,4.5) {Empty\strut};
        \node[block] at (6.6,4.5) {Filled\strut};
        \draw[white, thick] (4.5,0.4) rectangle (7,1.8);
        \node[block] at (6.8,0.2) {Test block};
	\end{tikzpicture}
	\caption{\textbf{Top:} view inside the inner vessel from ELP 0.01 Lux webcam. \textbf{Bottom:} view from Xlayer USB Webcam. The picture shows the open-topped vessel empty on the left and after filling the full \qty{300}{kg} of xenon on the right, where the liquid xenon fully covers the stainless steel test block.}
	
	\label{fig:highres_pic}
\end{figure}
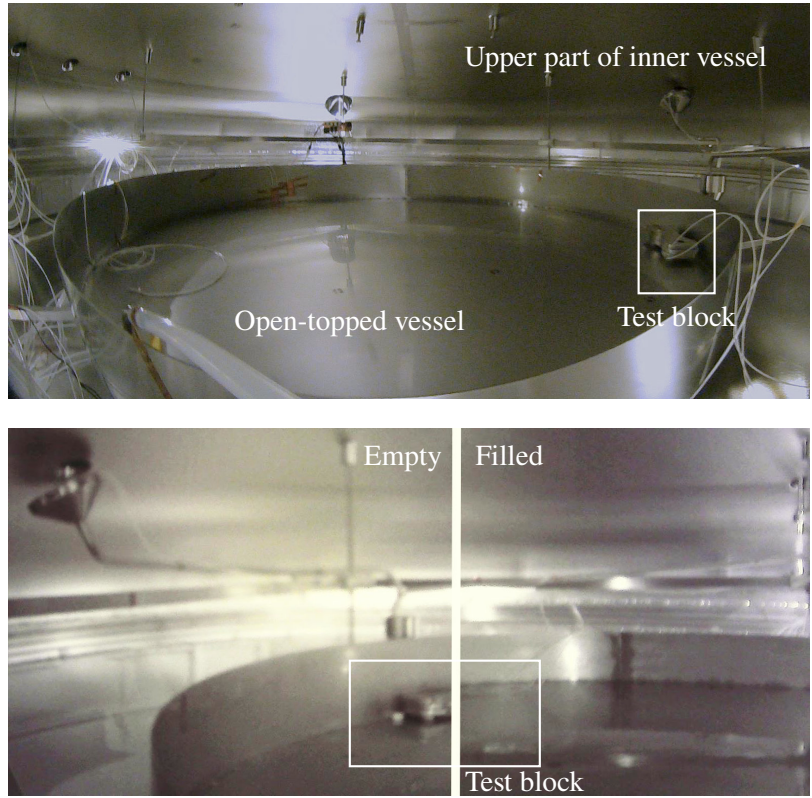

Cameras are installed in the inner vessel to observe components being tested and monitor the liquid xenon.
Tests of various low-cost USB-powered off-the-shelf webcams, including an ELP 0.01 Lux and an XLayer USB webcam showed that operation is possible in cold nitrogen and argon gas down to around \SI{-80}{\celsius}.
Stable operation in xenon at \SI{-100}{\celsius} can be achieved by heating the cameras with about \SI{5}{W} each.
After removing the plastic housing, Fischer Elektronik WLK30 thermally conductive epoxy is used to attach the camera to a \SI{1}{mm} copper plate which is heated with a polyimide heating tape.
So far we have used USB 2.0 to communicate with the cameras, either via an off-the-shelf MDC USB~2.0 feedthrough or a standard multi-pin vacuum feedthrough.
Consumer-grade USB hubs, heated in the same way as the cameras, can be used to provide power and data connection to at least two cameras.
Two views inside the inner vessel from an ELP 0.01 Lux and an Xlayer USB webcam are shown in \autoref{fig:highres_pic}.
For illumination, white \SI{3}{V} LEDs were installed, producing a total of around \SI{2}{W} of heat.

\subsection{High-voltage supply}

A Heinzinger PNC~60000-3 high-voltage supply allows components such as electrodes to be tested.
It can deliver up to \qty{3}{\uA} at \qty{\pm60}{kV}.
The high voltage is carried into the inner vessel using an MDC HV40-1S-C40 ceramic DN40 CF feedthrough, rated to \qty{40}{kV}.
On the inside there is a custom extension consisting of PTFE-insulated copper wire, to the end of which a test component can be connected.

\section{Commissioning}
\label{sec:commissioning}

In this section, we describe the commissioning of the facility with \qty{300}{kg} of xenon between October 2022 and April 2023.
An initial test with liquid argon verified the performance of the inner vessel's seal at temperatures below \qty{-100}{\celsius}. 
Following this, \qty{50}{\kg} of xenon was used to demonstrate the functionality of the xenon recuperation system.
Stable operation was then verified with \qty{300}{\kg} of xenon over the course of two weeks.
All these tests were performed without reopening the inner vessel.

An additional open-topped vessel with a smaller diameter can be placed within the inner vessel to submerge test components within deeper liquid xenon.
For the results described here, a cylindrical vessel with an inner diameter of \qty{1460}{mm} and a height of \qty{150}{mm} was used, as seen in \autoref{fig:highres_pic}.
This was constructed from \qty{2}{mm} stainless steel.
Liquefied xenon from the thermosyphon is directed into this vessel at the bottom of the side wall.

\subsection{Xenon filling and recuperation}
We tested the filling, liquefaction and recovery processes using \qty{50}{kg} xenon at \qty{1.2}{bar}, which results in about \qty{10}{kg} of liquid.
Liquefaction started at a temperature of \qty{-109}{\celsius} at \qty{1.0}{bar}.
After one week of stable operation, the \qty{50}{kg} of xenon was recovered via the automated emergency line (see \autoref{sec:storagesystem}) by means of cryogenic pumping into one of the cold bottles.

\begin{figure}
    \centering
    \includegraphics{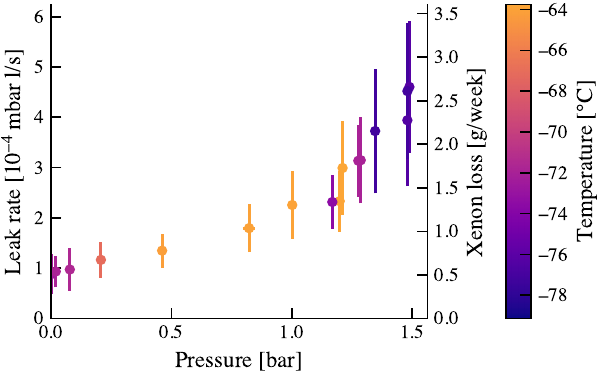}
    \caption{The leak rate and the corresponding mass loss rate depend on the xenon pressure in the inner vessel. Points are coloured according to the flange temperature of between \qty{-79}{\celsius} and \qty{-64}{\celsius}. The non-zero leak rate even at low pressure is the result of a leak from the pipes connecting the external nitrogen cooling pads; this was fixed before the last three points above \qty{1.3}{bar}.}
    \label{fig:leakrates}
\end{figure}

While filling, we regularly measured the inner vessel seal's leak rate by directing the outlet from the outer vessel's turbomolecular pump into an evacuated \qty{37}{L} vessel and monitoring the rate of pressure rise.
This rate-of-rise measurement with an external vessel is much more sensitive than using the full \qty{6.5}{m^3} volume of the insulation vacuum and doesn't disturb the stable operation of the system.
In equilibrium conditions and ignoring outgassing, the flow through the turbomolecular pump is the same as the leak rate from the inner vessel.
The results of these measurements are shown in \autoref{fig:leakrates}. At a xenon pressure of \qty{1.2}{bar} and a flange temperature of \qty{-64}{\celsius}, the leak rate was \qty{2.7(11)e-4}{mbar.L/s} corresponding to an expected xenon loss of \qty{1.6(0.7)}{\gram} per week.
Due to the thermal gradient in the inner cryostat vessel's wall, this is roughly the temperature of the flange when in equilibrium with liquid xenon in the cryostat.
Prior to filling xenon, we measured the leak rate at a flange temperature of \qty{-126}{\celsius}, below the operational conditions.
With an argon pressure of \qty{1.8}{bar} at this temperature, the leak rate was \qty{3(0.1)e-2}{mbar.L/s}.
Flange deformation, predicted by finite element analysis, results in a non-linear relationship between inner vessel pressure and leak rate.
The difference in leak rate from that measured with argon may be explained by the lower pressure, higher flange temperature and larger atomic size.

\subsection{Final commissioning run}
\begin{figure}
    \centering
    \includegraphics[width = \textwidth]{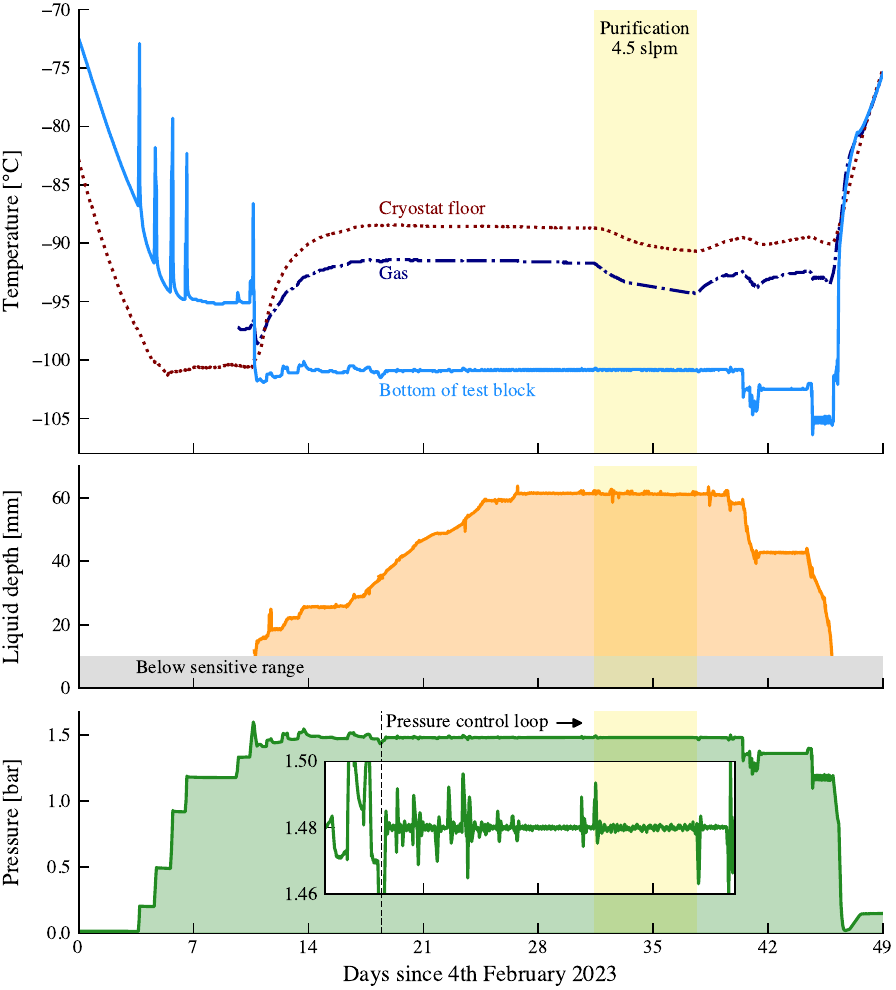}
    \caption{Operating conditions during the final \qty{300}{kg} commissioning run. \textbf{Top:} temperature profile within the inner cryostat vessel. Three PT100 sensors monitor the floor of the inner cryostat vessel, the stainless steel test block in the open-topped vessel and the gaseous xenon. Spikes in the test-block temperature during the initial cool-down correspond to times when additional xenon was filled into the inner vessel.
    \textbf{Middle:} liquid xenon depth, measured using level meters within the open-topped vessel. Depths below \qty{10}{mm} are outside the sensitive range of the level meters. \textbf{Bottom:} inner cryostat xenon pressure, with the inset showing a zoom during the time period where a control loop was implemented to stabilise the pressure. Residual fluctuations coincide with adjustments to the filling or purification flow. The yellow band visible across all subplots indicates the period when xenon was circulated through the purification system.}
    \label{fig:run_overview}
\end{figure}

We performed the final commissioning of the facility using \qty{300}{kg} of xenon, to demonstrate its stable operation.
This operation took about two months, including about one week to cool the inner cryostat vessel starting from \qty{-65}{\celsius}, three weeks to fill xenon, two weeks of stable operation including one week circulating xenon through the gas purifier, and two weeks to recover the xenon.
The temperature within the inner vessel, liquid xenon depth within the open-topped vessel and gas pressure during the full period are shown in~\autoref{fig:run_overview}.

When testing delicate experimental components using the PANCAKE facility, it may be necessary to control temperature gradients to avoid stress due to thermal contraction.
To estimate the temperature differences that might be produced in such a component, a \qty{20}{mm} thick stainless steel test block was placed within the open-topped vessel.
This was monitored using PT100 temperature sensors in contact with the top and bottom surfaces of the block.
Before filling any xenon, the inner vessel was pre-cooled to \qty{-100}{\celsius} using only the external pads.
Gaseous xenon was then filled into the inner cryostat vessel to a sufficient pressure for liquefaction to start at the thermosyphon's copper condenser.
The external cooling pads were then switched off and only the thermosyphon was used to cool.
This procedure is designed to reduce the temperature gradients produced within test components when liquid xenon starts to be collected in the open-topped vessel.
While filling, a maximum difference of \qty{10}{\celsius} was observed between the block and the gas within the cryostat.

Over a period of three weeks, \qty{300}{\kg} of xenon was filled at a typical flow of about 5 standard litres per minute (slpm). While filling, an inner cryostat pressure of \qty{1.48}{bar} was maintained by controlling the temperature of the xenon condenser.
Part-way through filling, a control loop was implemented in the SCADA system to adjust the xenon condenser temperature based on the xenon pressure in the inner vessel. This resulted in pressure variations smaller than \qty{0.02}{bar}, as seen in the lower panel's inset of \autoref{fig:run_overview}.
When not filling, the control loop maintains the pressure within \qty{1}{mbar} of the target.
During filling, the leak rate was regularly measured as described above, reaching \qty{4(2)e-4}{mbar.L/s} at \qty{1.48}{bar} and a flange temperature of \qty{-77}{\celsius} (see \autoref{fig:leakrates}).
The leak rate does not seem to depend strongly on the temperature, which first increased and then decreased over the course of the leak rate measurements due to the procedure of first cooling and then filling gaseous xenon before finally liquefaction began.

Two of the capacitive level meters described in \ref{sec:instrumentation} within the open-topped vessel monitored the liquid xenon depth, which reached \qty{61(1)}{mm} when all \qty{300}{kg} of xenon was filled.
These were installed partially overlapping in order to be sensitive to a greater range of depths between \qty{10}{mm} and \qty{75}{mm}.
Directing liquefied xenon into the open-topped vessel resulted in the floor of the inner vessel stabilising at a temperature above the xenon boiling point.
A third level meter installed in the inner vessel's bottom feedthrough was used to confirm the absence of condensed liquid xenon there while the open-topped vessel was full.

After filling and operating stably for four days, we started circulating xenon through the purification system.
Flows of up to \qty{15}{slpm} through the gas purifier were achieved when extracting liquid xenon through the heat exchanger, corresponding to a \qty{2.4}{day} turnover time for the \qty{300}{kg} in the cryostat.
The additional cooling power needed was less than \qty{10}{W}, indicating a heat exchanger efficiency of better than 90\%. 
During this operation, the inner cryostat floor began to cool down, as can be seen in \autoref{fig:run_overview}.
This can be explained by xenon condensing on the outside of the pipe used to carry liquid xenon from the open-topped vessel to the heat exchanger.
The condensed xenon was able to drip directly onto the inner cryostat vessel's floor, where its immediate evaporation results in cooling. 

The xenon was recovered into the storage system over seven days, leaving less than \qty{25}{mbar} in the inner vessel.
The total mass of xenon filled was \qty{301.5(10)}{kg}, determined from the change in weight of the warm and cold storage.
This agrees with the integrated mass flow when filling of \qty{292(15)}{kg}.

\section{Summary}
\label{sec:summary}

The PANCAKE facility is the world's largest liquid-xenon test facility, designed to test components for future liquid-xenon-based experiments.
The core of the facility is a vacuum-insulated stainless-steel cryostat with an internal diameter of~\qty{2.75}{m} and a flat floor.
Using two independent liquid-nitrogen cooling systems, the inner cryostat vessel can be cooled down and filled in around six weeks.
Xenon storage and purification systems allow controlled filling, cleaning and recovery of the xenon.
The facility is continuously monitored by a SCADA system, allowing critical procedures to be controlled remotely or automatised.
Following its construction, the facility was commissioned at the University of Freiburg with \qty{300}{\kg} of xenon.
Over six months, all major systems were tested and stably operated.

The platform is now operational and ready for upcoming test runs.
At present, \qty{380}{kg} xenon is available, allowing a liquid depth of \qty{18}{mm} across the \qty{2.75}{m} diameter of the inner vessel at \qty{2}{bar}.
Deeper liquid can be used on smaller diameters by placing an additional open-topped vessel within the inner cryostat vessel.

Both the inner cryostat vessel's flat floor and the open-topped vessel may be interesting concepts for future large-scale xenon-based rare-event detectors.
The flat floor saves several tonnes of xenon which would otherwise be needed below the active region.
However, the additional stiffening structure below our inner cryostat vessel requires roughly an additional \qty{1}{t} of stainless steel.
A similar stiffening structure in a rare-event detector could add to the background.
The open-topped vessel, also with a flat floor, is an alternative.
Since gaseous xenon is also present below the vessel, only the mass of xenon inside must be supported and not the gas pressure itself.
Such a vessel, or a similar gas-pressure-supported flat floor, could also be installed inside a pressure vessel with a standard torispherical head.
This significantly reduces the pressure on the flat floor and therefore also the amount of material needed.

By using PANCAKE, critical components of future liquid-xenon detectors can be tested directly in the environment they will be exposed to in a future dark-matter detector.
An example is the electrodes, which have historically proven to be challenging.
As well testing their mechanical stability during the cooling-down and warming-up procedures, the PANCAKE facility provides the possibility of applying high voltages to test their electric performance.
Real-time visual feedback is available using cameras installed within the inner cryostat vessel.
We also foresee installing photomultiplier tubes (PMTs) in the inner vessel in order to be sensitive to lower levels of light emission.

\section*{Acknowledgments}
This work was supported by the large-scale infrastructure grant ``DARWIN Demonstrator'' funded jointly by the DFG (grant number INST~39/1095-1 FUGG) and the SI-BW of the state of Baden-Württemberg, the European Research Council (ERC) grant number~724320 (ULTIMATE) and the German  Ministry  for  Education and  Research (BMBF) grant number~05A17VF1. We gratefully acknowledge the continuous support from the mechanical and electrical workshops of the Institute of Physics at the University of Freiburg.

%\bibliography{main.bib}{}

\begin{thebibliography}{10}

\bibitem{XENON:2023cxc}
{\scshape XENON} collaboration, E.~Aprile et~al., \emph{First dark matter search with nuclear recoils from the {XENONnT} experiment}, \href{http://dx.doi.org/10.1103/PhysRevLett.131.041003}{\emph{Phys. Rev. Lett.} {\bfseries 131} (2023) 041003}, [\href{https://arxiv.org/abs/2303.14729}{{\ttfamily 2303.14729}}].

\bibitem{LZ:2022lsv}
{\scshape LZ} collaboration, J.~Aalbers et~al., \emph{First dark matter search results from the {LUX-ZEPLIN (LZ)} experiment}, \href{http://dx.doi.org/10.1103/PhysRevLett.131.041002}{\emph{Phys. Rev. Lett.} {\bfseries 131} (2023) 041002}, [\href{https://arxiv.org/abs/2207.03764}{{\ttfamily 2207.03764}}].

\bibitem{PandaX-4T:2021bab}
{\scshape PandaX-4T} collaboration, Y.~Meng et~al., \emph{Dark matter search results from the {PandaX-4T} commissioning run}, \href{http://dx.doi.org/10.1103/PhysRevLett.127.261802}{\emph{Phys. Rev. Lett.} {\bfseries 127} (2021) 261802}, [\href{https://arxiv.org/abs/2107.13438}{{\ttfamily 2107.13438}}].

\bibitem{NEXT:2023daz}
{\scshape NEXT} collaboration, P.~Novella et~al., \emph{Demonstration of neutrinoless double beta decay searches in gaseous xenon with {NEXT}}, \href{http://dx.doi.org/10.1007/JHEP09(2023)190}{\emph{JHEP} {\bfseries 09} (2023) 190}, [\href{https://arxiv.org/abs/2305.09435}{{\ttfamily 2305.09435}}].

\bibitem{EXO-200:2019rkq}
{\scshape EXO-200} collaboration, G.~Anton et~al., \emph{Search for neutrinoless double-$\beta$ decay with the complete {EXO-200} dataset}, \href{http://dx.doi.org/10.1103/PhysRevLett.123.161802}{\emph{Phys. Rev. Lett.} {\bfseries 123} (2019) 161802}, [\href{https://arxiv.org/abs/1906.02723}{{\ttfamily 1906.02723}}].

\bibitem{Schumann:2015cpa}
M.~Schumann et~al., \emph{{Dark matter sensitivity of multi-ton liquid xenon detectors}},\\ \href{http://dx.doi.org/10.1088/1475-7516/2015/10/016}{\emph{JCAP} {\bfseries 10} (2015) 016}, [\href{https://arxiv.org/abs/1506.08309}{{\ttfamily 1506.08309}}].

\bibitem{nEXO:2017nam}
{\scshape nEXO} collaboration, J.~B. Albert et~al., \emph{Sensitivity and discovery potential of {nEXO} to neutrinoless double beta decay}, \href{http://dx.doi.org/10.1103/PhysRevC.97.065503}{\emph{Phys. Rev. C} {\bfseries 97} (2018) 065503}, [\href{https://arxiv.org/abs/1710.05075}{{\ttfamily 1710.05075}}].

\bibitem{NEXT:2020amj}
{\scshape NEXT} collaboration, C.~Adams et~al., \emph{Sensitivity of a tonne-scale {NEXT} detector for neutrinoless double beta decay searches}, \href{http://dx.doi.org/10.1007/JHEP08(2021)164}{\emph{JHEP} {\bfseries 2021} (2021) 164}, [\href{https://arxiv.org/abs/2005.06467}{{\ttfamily 2005.06467}}].

\bibitem{DARWIN:2016hyl}
{\scshape DARWIN} collaboration, J.~Aalbers et~al., \emph{{DARWIN}: towards the ultimate dark matter detector},\\ \href{http://dx.doi.org/10.1088/1475-7516/2016/11/017}{\emph{JCAP} {\bfseries 11} (2016) 017}, [\href{https://arxiv.org/abs/1606.07001}{{\ttfamily 1606.07001}}].

\bibitem{Aalbers:2022dzr}
J.~Aalbers et~al., \emph{A next-generation liquid xenon observatory for dark matter and neutrino physics},\\ \href{http://dx.doi.org/10.1088/1361-6471/ac841a}{\emph{J. Phys. G} {\bfseries 50} (2023) 013001}, [\href{https://arxiv.org/abs/2203.02309}{{\ttfamily 2203.02309}}].

\bibitem{Baudis:2021ipf}
L.~Baudis et~al., \emph{{Design and construction of Xenoscope \textemdash{} a full-scale vertical demonstrator for the DARWIN observatory}}, \href{http://dx.doi.org/10.1088/1748-0221/16/08/P08052}{\emph{JINST} {\bfseries 16} (2021) P08052}, [\href{https://arxiv.org/abs/2105.13829}{{\ttfamily 2105.13829}}].

\bibitem{Bradley:2012fsa}
A.~W. Bradley et~al., \emph{{LUX} cryogenics and circulation}, \href{http://dx.doi.org/10.1016/j.phpro.2012.03.734}{\emph{Phys. Procedia} {\bfseries 37} (2012) 1122--1130}.

\bibitem{Zappa:2016zsn}
P.~Zappa et~al., \emph{{A versatile and light-weight slow control system for small-scale applications}},\\ \href{http://dx.doi.org/10.1088/1748-0221/11/09/T09003}{\emph{JINST} {\bfseries 11} (2016) T09003}, [\href{https://arxiv.org/abs/1607.08189}{{\ttfamily 1607.08189}}].

\bibitem{Baur:2022sel}
D.~Baur et~al., \emph{{The XeBRA platform for liquid xenon time projection chamber development}},\\ \href{http://dx.doi.org/10.1088/1748-0221/18/02/T02004}{\emph{JINST} {\bfseries 18} (2023) T02004}, [\href{https://arxiv.org/abs/2208.14815}{{\ttfamily 2208.14815}}].

\bibitem{Garcia:2022jdt}
D. Ramírez García et~al., \emph{{GeMSE: a low-background facility for gamma-spectrometry at moderate rock overburden}}, \href{http://dx.doi.org/10.1088/1748-0221/17/04/P04005}{\emph{JINST} {\bfseries 17} (2022) P04005}, [\href{https://arxiv.org/abs/2202.06540}{{\ttfamily 2202.06540}}].

\bibitem{XENON:2017lvq}
{\scshape XENON} collaboration, E.~Aprile et~al., \emph{The {XENON1T} dark matter experiment},\\ \href{http://dx.doi.org/10.1140/epjc/s10052-017-5326-3}{\emph{Eur. Phys. J. C} {\bfseries 77} (2017) 881}, [\href{https://arxiv.org/abs/1708.07051}{{\ttfamily 1708.07051}}].

\bibitem{Toschi_thesis}
F.~Toschi, \emph{Design of the field cage and charge response of the XENONnT dark matter experiment}.
\newblock PhD thesis, \href{http://dx.doi.org/10.6094/UNIFR/234323}{\emph{Albert-Ludwigs-Universität Freiburg}, Freiburg, 2022}.

\bibitem{Geis_thesis}
C.~W. Geis, \emph{The XENON1T water Cherenkov muon veto system and commissioning of the XENON1T dark matter experiment}.
\newblock PhD thesis, \href{http://dx.doi.org/10.25358/openscience-2175}{\emph{Johannes Gutenberg-Universität Mainz}, Mainz, 2018}.

\end{thebibliography}
%\bibliographystyle{JHEP}

\providecommand{\href}[2]{#2}\begingroup\raggedright\endgroup

\end{document}